\documentclass[letter]{jpconf}
\usepackage{graphicx}
\newcommand{\be}{\begin{equation}}
\newcommand{\ee}{\end{equation}}
\newcommand{\als}{\alpha_s}
\newcommand{\nn}{\nonumber}
\begin{document}
\title{Static quark correlators and quarkonium properties at non-zero temperature}

\author{A. Bazavov and P. Petreczky}

\address{Physics Department, Brookhaven National Laboratory, Upton, NY 11973}

\ead{bazavov@bnl.gov, petreczk@bnl.gov}

\begin{abstract}
We discuss different static quark correlators, including Wilson loops
in 2+1 flavor QCD at non-zero temperature
and their relation to in-medium quarkonium properties. We present lattice
results on static correlation functions obtained with highly
improved staggered fermion action and their implications for potential models.
\end{abstract}

\section{Introduction}
At sufficiently high temperatures strongly interacting matter undergoes a transition to
a new state, often called quark-gluon plasma 
that is characterized by chiral symmetry restoration and color screening (see e.g.
Ref. \cite{Petreczky:2012rq} for a current review).
Experimentally such state of matter can be studied in relativistic heavy collisions.
There was considerable interest in the properties and the fate of
heavy quarkonium states at finite temperature since the famous conjecture 
by Matsui and Satz \cite{Matsui:1986dk}. It has been argued that color screening in
medium will lead to  quarkonium dissociation above deconfinement, which in turn
can signal quark-gluon plasma formation in heavy ion collisions. The basic
assumption behind the conjecture by Matsui and Satz was the fact that 
medium effects can be understood in terms of a temperature-dependent 
heavy-quark potential. Color screening makes the potential 
exponentially suppressed at distances larger than the Debye radius and 
therefore it cannot bind the heavy quark and anti-quark once the temperature is
sufficiently high.
Based on this idea 
potential models at finite temperature with different temperature dependent
potentials have been used over the last two decades to study quarkonium properties at finite
temperature (see e.g. Refs. \cite{Mocsy:2008eg,Bazavov:2009us} for recent reviews). 
It is not clear if and to what extent medium effects on quarkonium binding can be
encoded in a temperature dependent potential. Effective field theory
approach, namely the so-called thermal pNRQCD, can provide an answer to
this question \cite{Brambilla:2008cx}. The notion of the potential can
be defined using EFT approach both at zero and non-zero temperature.
Thermal pNRQCD that will be discussed in the next 
section is based on the weak-coupling techniques. 
To understand the non-perturbative aspects of color screening 
as well as to test the reliability of the weak-coupling approach lattice calculations 
of the correlation functions of static quarks  are needed. The correlation functions
of static quarks that propagate around the periodic time direction $\tau = 1/T$ are related to the free
energy of a static quark anti-quark pair. 
We will see that pNRQCD is a useful tool in understanding
the temperature dependence of the static correlators. We also consider Wilson loops
evaluated at time extent $\tau<1/T$. They are naturally related to the static energy
at non-zero temperature.  

In principle, it is possible to study the problem of quarkonium dissolution without any use
of potential models. In-medium properties of different quarkonium states and/or their
dissolution are encoded in spectral functions. Spectral functions are related to Euclidean 
meson correlation functions which can be calculated on the lattice.
Reconstruction of the spectral functions from the lattice meson correlators turns out to be
very difficult, and the corresponding results 
remain inconclusive. We will discuss the calculation of the spectral functions
using potential models in the light of lattice calculations of Wilson loops.

\section{pNRQCD at finite temperature}
\label{sec_pnrqcd}

There are different scales in the heavy quark bound state problem related to the heavy quark mass $m$,
the inverse size $\sim m v \sim 1/r $ and the binding energy $~m v^2 \sim \alpha_s/r$. Here $v$ is the 
typical heavy quark velocity
in the bound state and is considered to be a small parameter.
Therefore it is possible to derive a sequence of effective field theories using this 
separation of scales (see Refs. \cite{Brambilla:2004jw,Brambilla:2010cs} for  recent reviews). 
Integrating out modes at the highest energy scale $\sim m$ leads to
an effective field theory called non-relativistic QCD or NRQCD, where the pair creation of heavy quarks is
suppressed by powers of the inverse mass and the heavy quarks are described by non-relativistic Pauli 
spinors \cite{Caswell:1985ui}.
At the next step, when the large scale related to the inverse size is integrated out, the potential NRQCD
or pNRQCD appears. In this effective theory the dynamical fields include the singlet 
$\rm S(r,R)$ and octet $\rm O(r,R)$ fields 
corresponding to the heavy quark anti-quark pair in singlet and octet states respectively, 
as well as light quarks and gluon fields
at the lowest scale $\sim mv^2$. The Lagrangian of this effective field theory has the form
\begin{eqnarray}
{\cal L} =
&&
- \frac{1}{4} F^a_{\mu \nu} F^{a\,\mu \nu}
+ \sum_{i=1}^{n_f}\bar{q}_i\,iD\!\!\!\!/\,q_i
+ \int d^3r \; {\rm Tr} \,
\Biggl\{ {\rm S}^\dagger \left[ i\partial_0 + \frac{\nabla_r^2}{m}-V_s(r) \right] {\rm S}\nonumber\\
&&
+ {\rm O}^\dagger \left[ iD_0 + \frac{\nabla_r^2}{m}- V_o(r) \right] {\rm O} \Biggr\}
+ V_A\, {\rm Tr} \left\{  {\rm O}^\dagger {\vec r} \cdot g{\vec E} \,{\rm S}
+ {\rm S}^\dagger {\vec r} \cdot g{\vec E} \,{\rm O} \right\}
\nonumber\\
&&
+ \frac{V_B}{2} {\rm Tr} \left\{  {\rm O}^\dagger {\vec r} \cdot g{\vec E} \, {\rm O}
+ {\rm O}^\dagger {\rm O} {\vec r} \cdot g{\vec E}  \right\}  + \dots\;.
\label{pNRQCD}
\end{eqnarray}
Here the dots correspond to terms which are higher order in the multipole expansion \cite{Brambilla:2004jw}.
The relative distance $r$ between the heavy quark and anti-quark plays a role of a label, the light
quark and gluon fields depend only on the center-of-mass coordinate $R$. The singlet $V_s(r)$ and octet $V_o(r)$ 
heavy quark potentials
appear as matching coefficients in the Lagrangian of the effective field theory
and therefore can be rigorously defined in QCD at any order of the perturbative expansion.
At leading order 
\be
V_s(r)=-\frac{4}{3} \frac{\alpha_s}{r},~V_o(r)=\frac{1}{6}\frac{\alpha_s}{r}
\ee
and $V_A=V_B=1$.
One can generalize this approach to finite temperature. However, the presence of additional 
scales makes the analysis more complicated \cite{Brambilla:2008cx}. The effective Lagrangian will have the same form as above,
but the matching coefficients may be temperature-dependent. In the weak coupling regime there are three different thermal scales:
$T$, $g T$ and $g^2 T$. The calculations of the matching coefficients depend on the relation of these thermal scales to
the heavy quark bound-state scales \cite{Brambilla:2008cx}. To simplify the analysis the static approximation has been used, in which
case the scale $m v$ is replaced by the inverse distance $1/r$ between the static quark and anti-quark. The binding energy
in the static limit becomes $V_o-V_s \simeq N \alpha_s/(2 r)$. When the binding energy is larger than the temperature the 
derivation of pNRQCD proceeds
in the same way as at zero temperature and there is no medium modifications of 
the heavy quark potential \cite{Brambilla:2008cx}. 
But bound state
properties will be affected by the medium through interactions with ultra-soft gluons, in particular, 
the binding energy will be reduced
and a finite thermal width will appear due to medium induced singlet-octet transitions arising from the dipole interactions in
the pNRQCD Lagrangian \cite{Brambilla:2008cx} (c.f. Eq. (\ref{pNRQCD})).
When the binding energy is smaller than one of the thermal scales the singlet
and octet potential will be temperature-dependent and will acquire an imaginary part \cite{Brambilla:2008cx}. 
The imaginary part of the potential arises because of the singlet-octet transitions induced by the dipole vertex as well as
due to the Landau damping in the plasma, i.e. scattering of the gluons with space-like momentum off the thermal excitations in
the plasma. 
In general, the thermal corrections
to the potential go like $(r T)^n$ and $(m_D r)^n$ \cite{Brambilla:2008cx}, where $m_D$ denotes the Debye mass. 
Only for distances $r>1/m_D$ there is an exponential screening. 
In this region the singlet potential has a simple form
\begin{equation}
V_s(r)=
 -\frac{4}{3} \,\frac{\als}{r}\,e^{-m_Dr}
+ i \frac{4}{3}\,\als\, T\,\frac{2}{rm_D}\int_0^\infty dx \,\frac{\sin(m_Dr\,x)}{(x^2+1)^2}-\frac{4}{3}\, \als \left( m_D + i T \right),\nn\\
\label{Vs}
\end{equation}
The real part of the singlet potential coincides with the leading order 
result of the so-called singlet free energy \cite{Petreczky:2005bd}.
The imaginary part of the singlet potential in this limit has been first calculated in \cite{Laine:2006ns}.
For small distances the imaginary part vanishes, while at large distances it is twice the damping rate of a
heavy quark \cite{Pisarski:1993rf}. This fact was first noted in Ref. \cite{Beraudo:2007ky} for thermal QED.

The effective field theory at finite temperature
has been derived in the weak-coupling regime assuming the separation of 
different thermal scales as well as $\Lambda_{QCD}$.
In practice, the separation of these scales is not evident and 
one needs lattice techniques to test the approach. 
Lattice QCD is formulated in Euclidean time. 
Therefore
the next section  will be dedicated to the study of 
static quarks at finite temperature in Euclidean time formalism.

\section{Static meson correlators in Euclidean time formalism}
\label{sec_meson_cor}

Consider static (infinitely heavy) quarks. The position
of heavy quark anti-quark pair is fixed in space and propagation happens only
along the time direction. 
With respect to the color a static quark anti-quark ($Q\bar Q$) pair 
can be in a singlet or in an octet state. Therefore we can define the following
$Q \bar Q$ (meson) operators 
\begin{eqnarray}
  \label{Jmesstat}
  J(\vec{x},\vec{y};\tau)&=&
  \bar\psi(\vec{x},\tau)U(\vec{x},\vec{y};\tau)\psi(\vec{y},\tau),\\
  \label{Jamesstat}
  J^a(\vec{x},\vec{y};\tau)&=&
  \bar\psi(\vec{x},\tau)U(\vec{x},\vec{x}_0;\tau)
  T^a U(\vec{x}_0,\vec{y};\tau)\psi(\vec{y},\tau),
\end{eqnarray}
for singlet and octet state respectively. Here $U(\vec{x},\vec{y};\tau)$ are the
spatial gauge transporters 
connecting $\vec{x}$ and $\vec{y}$, $\vec{x}_0$ is the coordinate
of the center of mass of the meson and $T^a$ are the $SU(3)$ group generators.
We consider the correlation function of singlet and octet meson
operators at maximal Euclidean time $\tau=1/T$:
\begin{eqnarray}
&
G_1(r,T,\tau=1/T)=\langle J(x,y,\tau=1/T) J^{\dagger}(x,y;0) \rangle,\\
&
G_8(r,T,\tau=1/T)=\frac{1}{8} \langle J^a(x,y,\tau=1/T) J^{a \dagger}(x,y;0) \rangle.
\end{eqnarray}
Integrating out the static quark fields $\psi$ and replacing
the quark propagators by  temporal Wilson lines $L(\vec{x})=\prod_{\tau=0}^{N_\tau-1} U_0(\vec{x},\tau)$
with $U_0(\vec{x},t)$ being the temporal links, 
we get  the following expression for the above correlators:
\begin{eqnarray}
\displaystyle
G_1(r,T) &=&\frac{1}{3} \langle {\rm Tr}\left[
L^{\dagger}(\vec{x}) U(\vec{x},\vec{y};0) 
L(\vec{y}) U^{\dagger}(\vec{x},\vec{y},1/T)\right]\rangle, \label{defg1}\\
\displaystyle
G_8(r,T)&=&\frac{1}{8}
\langle {\rm Tr} L^{\dagger}(x)  {\rm Tr} L(y) \rangle-\frac{1}{24} 
\langle {\rm Tr}\left[
L^{\dagger}(x) U(x,y;0) L(y) U^{\dagger}(x,y,1/T)\right] \rangle,
\label{defg3}\\
&&r=|\vec{x}-\vec{y}|. \nonumber
\end{eqnarray}
The correlators depend on the choice of the spatial transporters
$U(\vec{x},\vec{y};\tau)$. Typically, a straight line connecting points 
$\vec{x}$ and $\vec{y}$ is used as a path in the gauge transporters, 
i.e. one deals with time-like rectangular cyclic Wilson loops.
In the special gauge, where $U(\vec{x},\vec{y};\tau)=1$ the above
correlators give the standard definition of the singlet and
octet free energies of a static $Q\bar Q$ pair 
\begin{eqnarray}
\displaystyle
\exp(-F_1(r,T)/T)&=&
\frac{1}{3} \langle {\rm Tr}[L^{\dagger}(x)  L(y)]\rangle,\label{F1def}\\
\displaystyle
\exp(-F_8(r,T)/T)&=&
\frac{1}{8}
\langle {\rm Tr} L^{\dagger}(x)  {\rm Tr} L(y) \rangle
-\frac{1}{24} \langle {\rm Tr}
\left[L^{\dagger}(x)  L(y)\right]\rangle.\label{Fadef}
\end{eqnarray}
One can also fix the Coulomb gauge and define the interpolating meson operators without
the spatial transporters, and use the above expression to define the singlet and octet correlators. 
\begin{figure}
\includegraphics[width=7cm]{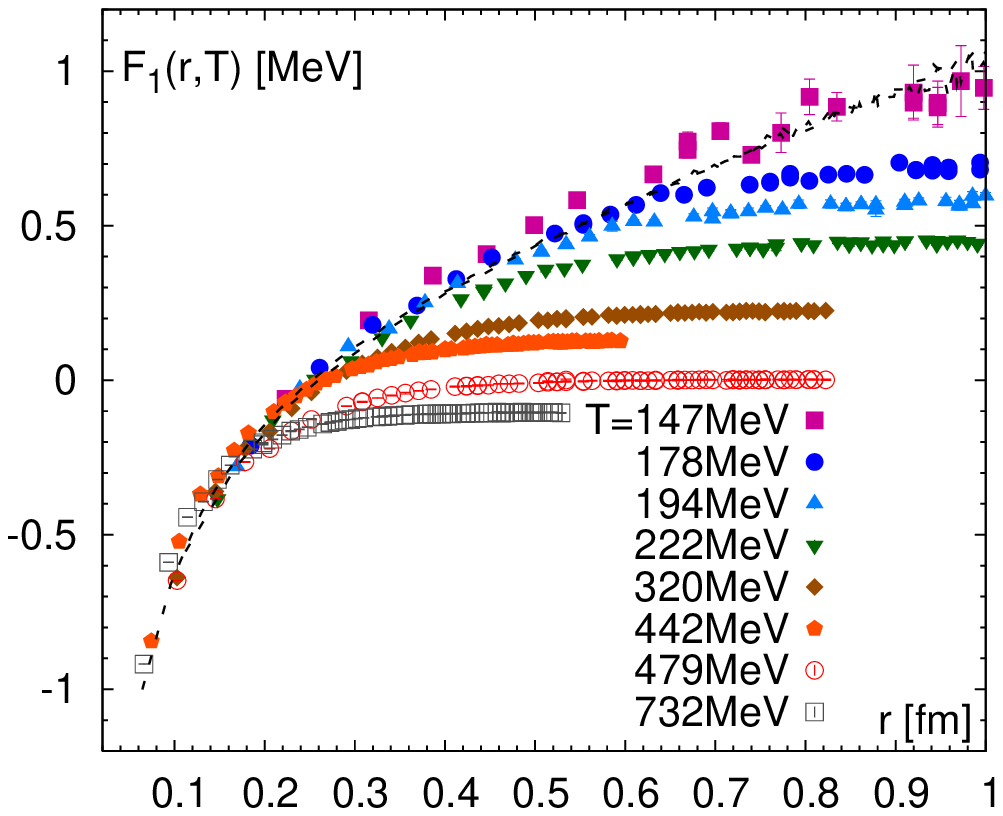}
\includegraphics[width=8cm]{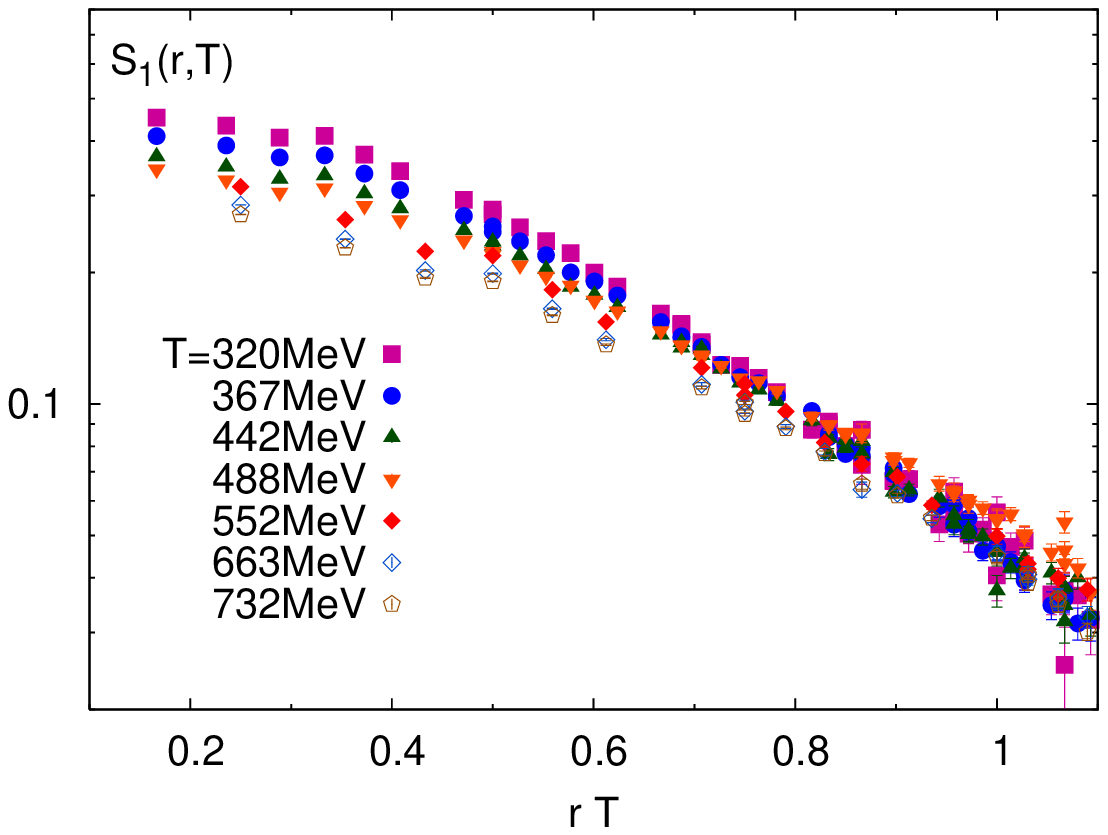}
\vspace*{-0.3cm}
\caption{The singlet free energy (left) and the screening function (right) as 
function of the distance $r$ at different temperatures calculated with the HISQ action.}
\label{fig:f1}
\vspace*{-0.3cm}
\end{figure}

The correlator $G(r,T)=\frac{1}{9} \langle {\rm Tr} L^{\dagger}(x)  {\rm Tr} L(y) \rangle$  gives
the free energy $F(r,T)=-T \ln G(r,T)$ of static quark anti-quark pair separated by distance $r$ \cite{McLerran:1981pb}.
It can be expressed in terms of energy levels $E_n(r)$ of static quark anti-quark pair at $T=0$ \cite{Jahn:2004qr}
\be
G(r,T)=\sum_{n=1}^{\infty} e^{-E_n(r)/T}.
\label{g}
\ee
It is tempting to rewrite Eq. (\ref{defg3}) or Eq. (\ref{Fadef}) as
\be
\exp(-F(r,T)/T)=\frac{1}{9} \exp(-F_1(r,T)/T)+\frac{8}{9} \exp(-F_8(r,T)/T)
\label{decomp}
\ee
and interpret this expression as the decomposition of the free energy of static $Q\bar Q$ pair into
singlet and octet contributions \cite{McLerran:1981pb,Gross:1980br,Nadkarni:1986cz,Nadkarni:1986as}.
This decomposition is intuitively very appealing and should be valid in perturbation theory.
However, it is problematic as $G_1(r,T)$ is path- or gauge- dependent. The problem
is also evident if one writes the spectral decomposition of $G_1(r,T)$ \cite{Jahn:2004qr}:
\be
 G_1(r,T)=\sum_{n=1}^{\infty} c_n(r) e^{-E_n(r,T)/T} \label{g1}.
\ee
The coefficients $c_n(r)$ are different from unity and are path- or gauge- dependent.
The EFT approach can help to resolve this puzzle. One can use pNRQCD also in Euclidean
time formulation \cite{Brambilla:2010xn} and study the Polyakov loop correlator in this
framework. The Polyakov loop correlator can be written in terms of correlation function
of singlet and octet fields \cite{Brambilla:2010xn}
\be
G(r,T)=Z_s(r) \langle S(r,\tau=1/T) S^{\dagger}(r,0)\rangle +Z_o(r) \langle O^a(r,\tau=1/T) O^{a \dagger}(r,0)\rangle.
\ee
For $r T \ll 1$ one can use the zero temperature version of pNRQCD where the singlet
and octet potentials are known up to 2-loop order. One can then show that $Z_s=Z_o=1/9$ and
thus in this limit the conjectured decomposition of the Polyakov loop correlator in terms
of singlet and octet contribution is justified \cite{Brambilla:2010xn}
\be
G(r,T)=\frac{1}{9} \exp(-V_s(r)/T)+ \frac{8}{9} \exp(-V_o(r)/T).
\label{decomp0}
\ee
The singlet and octet
contributions are gauge-independent in this framework. When the binding energy $E_{bin} \sim\als/r$
is the largest scale in the problem the free energy of $Q\bar Q$ pair is dominated by the singlet contribution
and is equal to the zero temperature potential \cite{Brambilla:2010xn} up to the term $T \ln 9$ coming from
the normalization constant. When the temperature is much larger
than the binding energy, i.e. $\als/(rT) \ll 1$ the exponentials in Eq. (\ref{decomp0}) can be expanded and
we get 
\be
F(r,T)=\frac{\als^2}{(r^2 T)}.
\ee
We see that despite no $T$-dependence of the potential in this limit, the free energy is strongly
temperature dependent and it is very different from the potential.
The complete next-to-leading order result can be found in Ref. \cite{Brambilla:2010xn}.

Similarly, for the singlet correlator one can write
\be
G_1(r,T)=\tilde Z_s(r) \langle S(r,\tau=1/T) S^{\dagger}(r,0)\rangle
\ee
At leading order $\tilde Z_s(r)=1$ and $F_1(r,T) \simeq V_s(r)$.
\begin{figure}
\includegraphics[width=7cm]{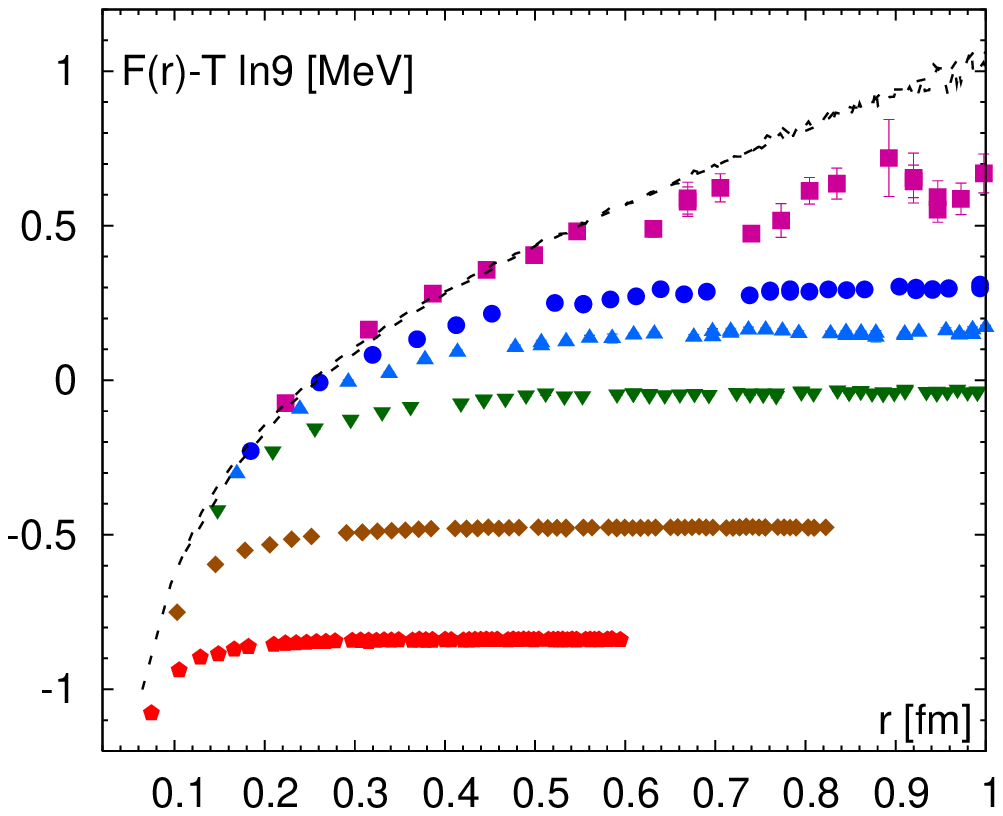}
\includegraphics[width=8cm]{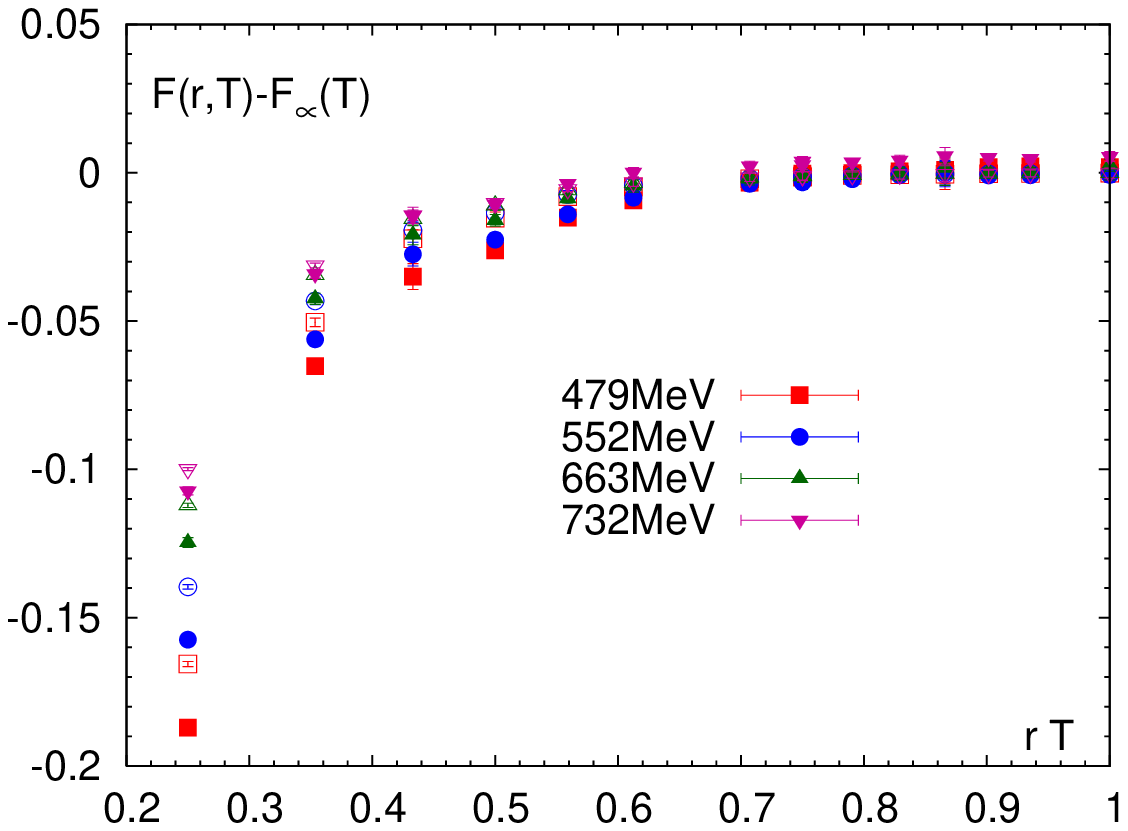}
\vspace*{-0.3cm}
\caption{The free energy of static $Q\bar Q$ pair  (left) and the 
difference  $F(r,T)-F_{\infty}(T)$ (right) calculated with HISQ action as 
function of the distance $r$ at different temperatures. In the right panel the filled symbols correspond
to the lattice data, while the open symbols correspond to the values reconstructed from
the singlet free energy.
The legend in the left panel is the same as in Fig. 1 (left).
}
\vspace*{-0.3cm}
\label{fig:f}
\end{figure}
At very high temperatures for $r \sim 1/m_D$
with $m_D=g T \sqrt{3/2}$ being the leading order Debye mass,  the singlet and octet correlators can be calculated in the 
hard thermal loop (HTL) approximation
\cite{Petreczky:2005bd}
\begin{equation}
F_1(r,T)=-\frac{4}{3} \frac{\alpha_s}{r} \exp(-m_D r)-\frac{4 \alpha_s m_D }{3},~~~
F_8(r,T)=\frac{1}{6} \frac{\alpha_s}{r} \exp(-m_D r)-\frac{4 \alpha_s m_D}{3}.
\label{f18p}
\end{equation}
The singlet and octet free energies are gauge-independent at this order. At large
distances the singlet and octet free energies approach a constant value $-\frac{4 \alpha_s m_D}{3}$.
This constant corresponds to the leading order result for the free energy of  two
isolated static quarks $F_{\infty}$, which has been also calculated to next-to-leading order 
\cite{Burnier:2009bk,Brambilla:2010xn}. 
The next-to-leading corrections are small and do not change the qualitative behavior
of $F_{\infty}(T)$ which decreases with increasing temperatures.
At leading order we have $(F_1(r,T)-F_{\infty}(T))/(F_8(r,T)-F_{\infty}(T))=-8$.

The free energy of static $Q\bar Q$ pair was calculated at leading order long time ago \cite{McLerran:1981pb,Gross:1980br,Nadkarni:1986cz}
\be 
F(r,T)=-\frac{1}{9} \frac{\alpha_s^2}{r^2} \exp(-2 m_D r)-\frac{4 \alpha_s m_D }{3}.
\ee
The above expression can also be obtained by inserting Eqs. (\ref{f18p}) into
Eq. (\ref{decomp}) and expanding the exponentials to order $\alpha_s^2$ thus confirming
the validity of the decomposition and the partial cancellation of the singlet and octet
contributions at leading order. The free energy was calculated at next-to-leading order for $r \simeq 1/m_D$
\cite{Nadkarni:1986cz} but
the decomposition into singlet and octet contributions was not shown.
Because of the partial cancellation of the singlet and octet contributions 
for $r \sim 1/m_D$ we expect that $F(r,T)-F_{\infty}(T) \ll F_1(r,T)-F_{\infty}(T)$ at sufficiently high temperatures. 
In summary, the partial cancellation of the singlet and octet contribution to the free energy happens both at short and
long distances and leads to its strong temperature dependence.

\section{Lattice results on the free energy and the singlet free energy}
We calculated Polyakov loop correlators as well as singlet correlators on the lattice
in 2+1 flavor QCD using Highly Improved Staggered Quark (HISQ) action \cite{Bazavov:2011nk} on $24^3 \times 6$ and $16^3 \times 4$ 
lattices. The strange quark mass $m_s$ was fixed to its physical value, while for the light quark masses
we used $m_l=m_s/20$ that corresponds to the pion mass of about $160$ MeV. The detailed choice of the lattice
parameters is discussed in Ref. \cite{Bazavov:2011nk}. To calculate the singlet free energy we used the
Coulomb gauge. The free energy and the singlet free energy have an additive divergent part that has to
be removed by adding a normalization constant determined from the zero temperature potential. 
We used the normalization constants from Ref.  \cite{Bazavov:2011nk}. 
The numerical results for the singlet free energy are shown in Fig. \ref{fig:f1}. At short distances the
singlet free energy agrees with the zero temperature potential calculated in Ref. \cite{Bazavov:2011nk},
while at large distances it approaches a constant value $F_{\infty}(T)$ equal to the excess free energy
of two isolated static quarks. As the temperature increases the deviation from the zero-temperature potential
shows up at shorter and shorter distances as the consequence of color screening. To explore the screening
behavior in Fig. \ref{fig:f1} we also show the combination $S(r,T)=r \cdot (F(r,T)-F_{\infty}(T))$ which
we call the screening function. The screening function 
should decay exponentially. We indeed observe the exponential decay of this quantity at distances larger
than $1/T$. Thus at high temperatures the behavior of the singlet free energy expected from the weak-coupling 
calculations seems to be confirmed by lattice QCD, at least qualitatively.
Let us also mention that at high temperatures
the behavior of the singlet free energy is similar to that observed in pure gauge theory \cite{Digal:2003jc,Kaczmarek:2002mc}.

In Fig. \ref{fig:f} we show our results for the free energy of static $Q \bar Q$ pair as function of the 
distance at different temperatures. At short distances and low temperatures the free energy is expected to be
dominated by the singlet contribution and we expect it to be equal to the zero temperature potential up to the
term $T \ln 9$ coming from the normalization, see the discussion in the previous section. 
Therefore in the figure the numerical results have been shifted by $-T \ln9$.
Indeed, for the smallest temperature and the shortest distances $F(r,T)-T \ln 9$ is equal to the zero temperature potential
shown as the dashed black line. At higher temperature $F(r,T)$ is very different from the zero-temperature potential.
At large distance the free energy approaches a constant value $F_{\infty}(T)$ that
decreases with increasing temperatures as expected (see discussions above). The temperature dependence of $F(r,T)$ is
much larger than that of the singlet free energy. This is presumably due to the partial cancellation of the singlet and octet
contribution discussed above. To verify this assertion we calculated $F(r,T)-F_{\infty}(T)$ using the numerical data for 
$F_1(r,T)-F_{\infty}(T)$ and the leading order relation  $(F_1(r,T)-F_{\infty}(T))/(F_8(r,T)-F_{\infty}(T))=-8$.
The corresponding results are shown in the right panel of Fig. \ref{fig:f}. As one can see from the figure
the numerical data for $F(r,T)$ are in reasonable agreement with the ones reconstructed from this procedure. The
reconstruction works better with increasing temperature. Thus the expected cancellation of the singlet and octet
contributions to the free energy of static $Q\bar Q$ pair seems to be confirmed by lattice calculations.

\begin{figure}
\hspace*{-0.7cm}
\includegraphics[width=5.9cm]{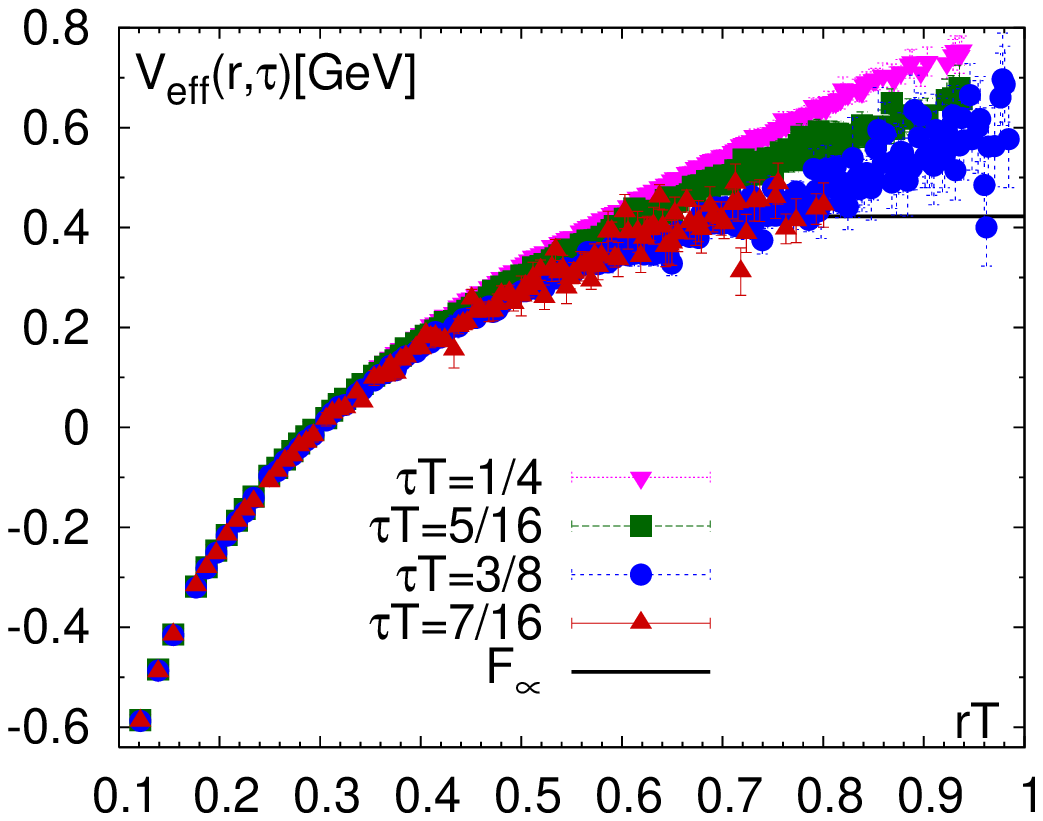}
\hspace*{-0.7cm}
\includegraphics[width=5.9cm]{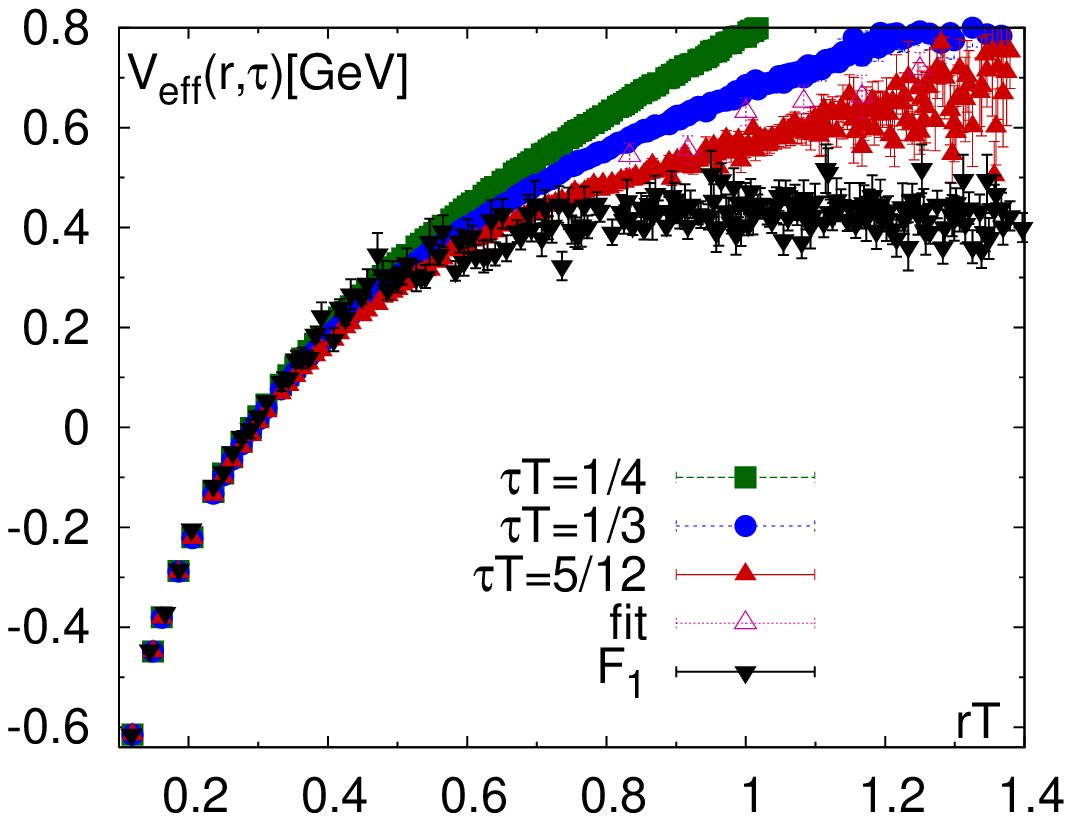}
\hspace*{-0.7cm}
\includegraphics[width=5.9cm]{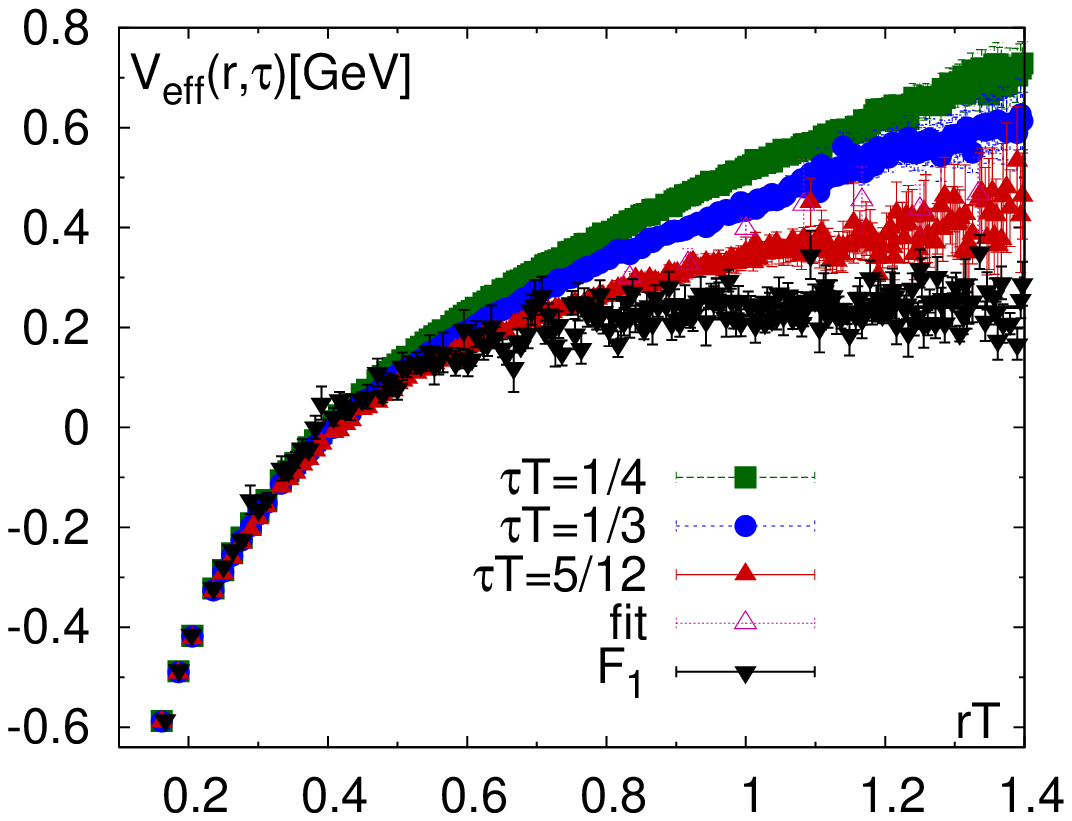}
\vspace*{-0.3truecm}
\caption{The effective potential $V_{eff}(r,\tau)$ as function of $r T$ calculated
for $48^3 \times 16$ lattice and $\beta=7.5$ (left), $48^3 \times 12$ lattice and $\beta=7.28$ (middle),
and $48^3 \times 12$ lattice and $\beta=7.5$ (right). The left, middle and right panels correspond
to temperatures of $225$ MeV, $249$ MeV and $300$ MeV respectively.}
\vspace*{-0.3truecm}
\label{fig:veff}
\end{figure}

\begin{figure}
\hspace*{-0.7cm}
\includegraphics[width=5.9cm]{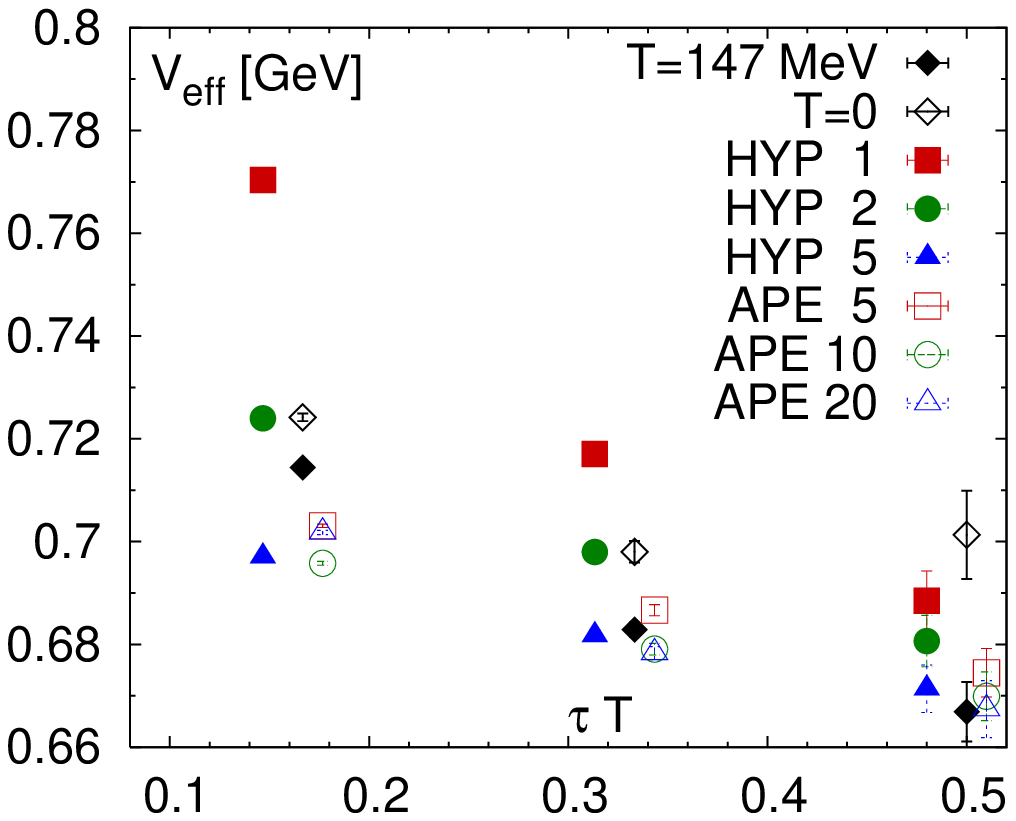}
\hspace*{-0.7cm}
\includegraphics[width=5.9cm]{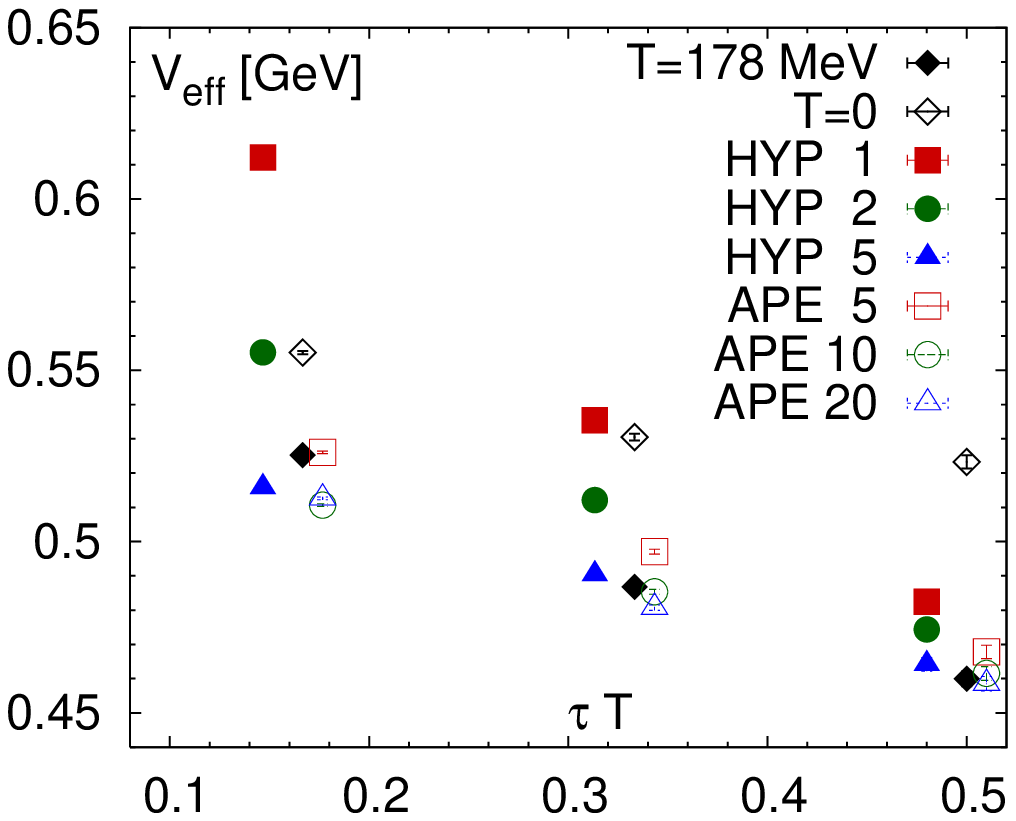}
\hspace*{-0.7cm}
\includegraphics[width=5.9cm]{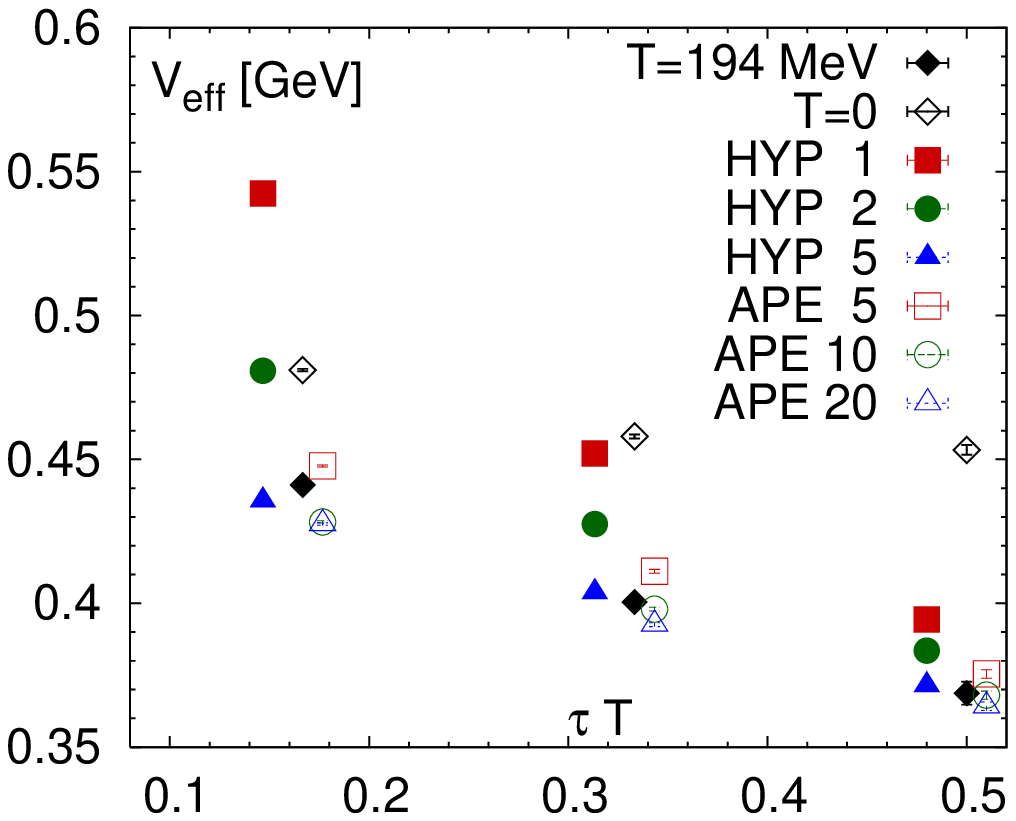}
\vspace*{-0.4cm}
\hspace*{-0.7cm}
\includegraphics[width=5.9cm]{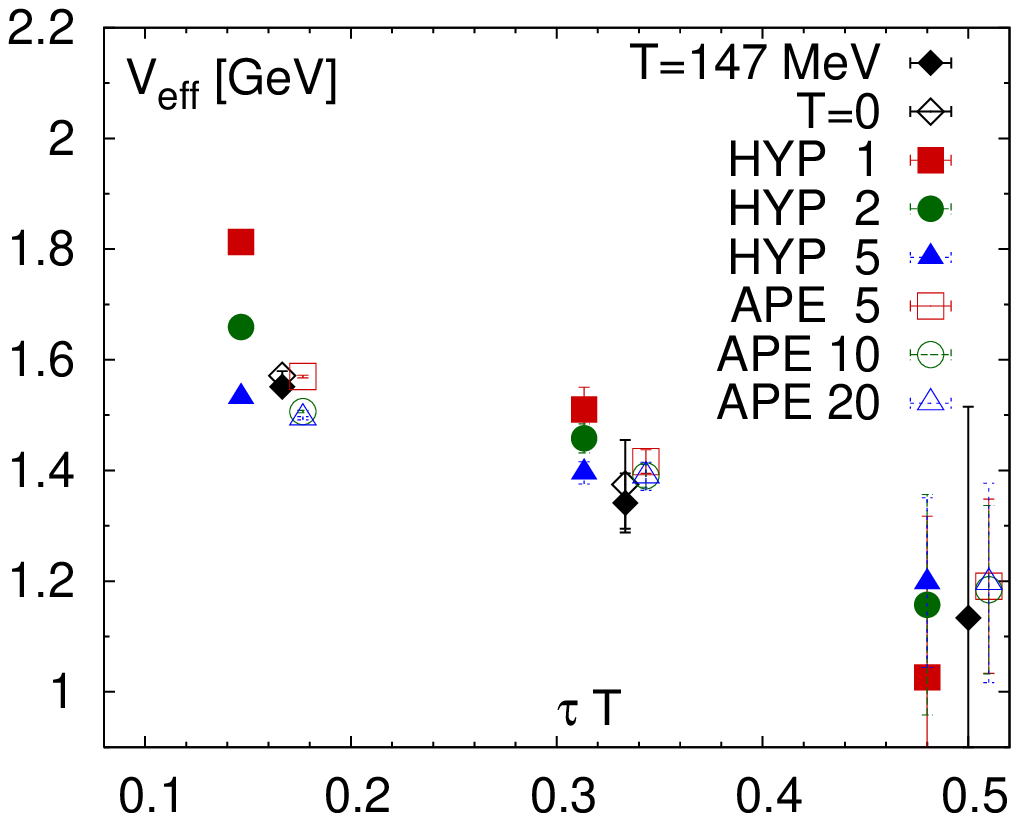}
\hspace*{-0.7cm}
\includegraphics[width=5.9cm]{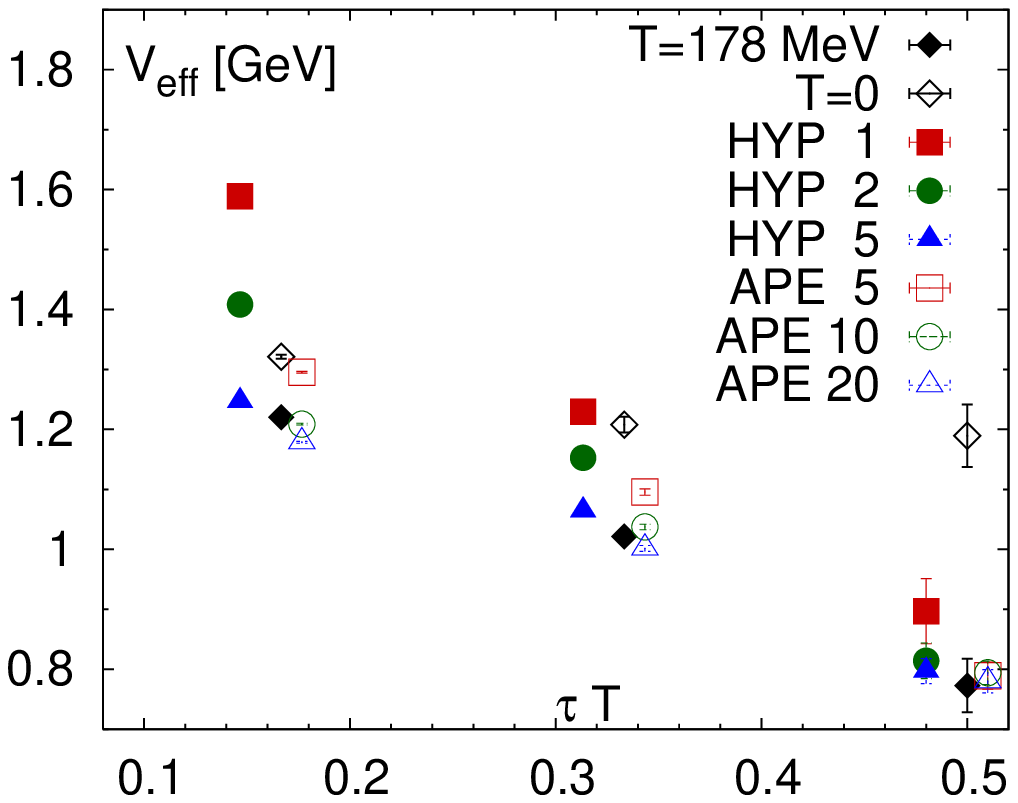}
\hspace*{-0.7cm}
\includegraphics[width=5.9cm]{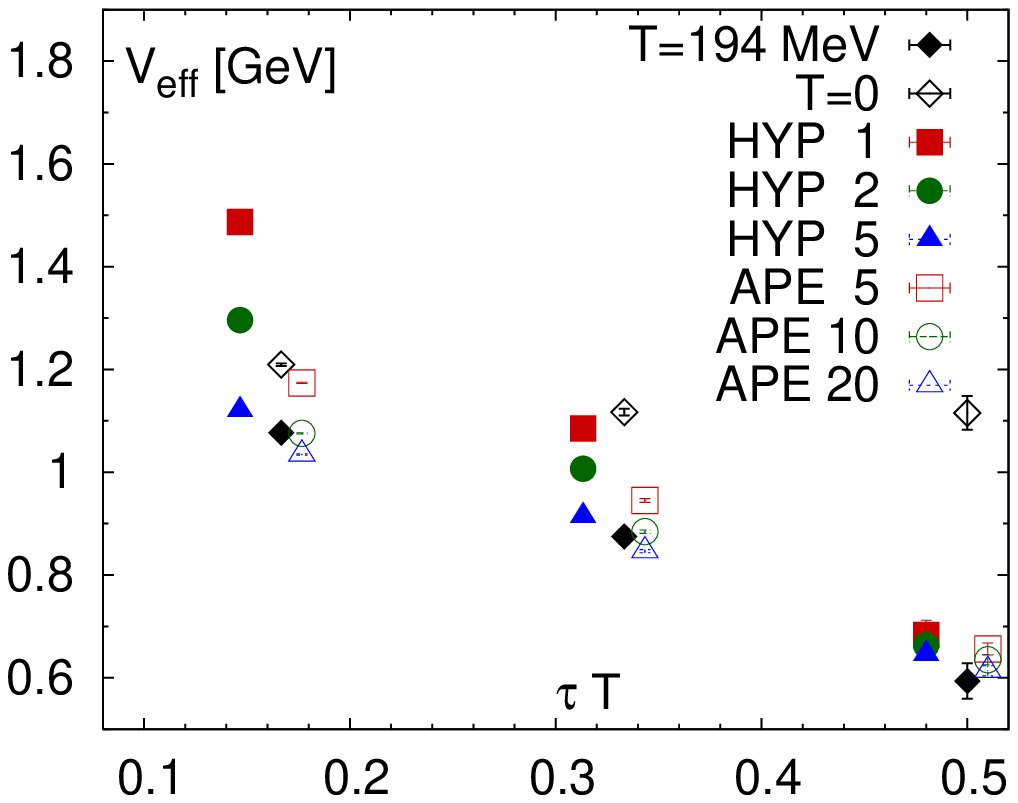}
\vspace*{-0.2truecm}
\caption{The effective potential $V_{eff}(r,\tau)$ as function of $\tau T$ calculated
on $24^3 \times 6$ lattices for three different temperatures. The upper panels correspond
to $rT =1/2$ while the lower panels correspond to $rT =1$. 
Filled and open diamonds correspond to the Coulomb gauge results at finite and zero temperature
respectively.}
\vspace*{-0.2truecm}
\label{fig:veff_t}
\end{figure}

\section{Wilson loops at non-zero temperature}
In the previous section we considered correlation function of static $Q\bar Q$ pair evaluated
at Euclidean time $t=1/T$. These correlators are related to the free energy of static $Q\bar Q$
pair. One can consider Wilson loops for Euclidean times $t < 1/T$ which have no obvious relation
to the free energy of a static $Q\bar Q$ pair. Wilson loops at non-zero temperature have been
first studied in Refs. \cite{Rothkopf:2009pk,Rothkopf:2011db} in connection with heavy quark potential at non-zero temperature 
and a spectral decomposition has been conjectured for the Wilson loops
\be
W(r,\tau)=\int_0^{\infty} \sigma(\omega,r,T) e^{-\omega \tau}.
\ee
At zero temperature the spectral function is proportional to sum of delta functions $\sigma(r,\omega)=\sum_n c_n \delta(E_n(r)-\omega)$
and thus the spectral decomposition is just the generalization of Eq. (\ref{g1}). At high temperatures the spectral function
will be proportional to sum of smeared delta functions and the position and the width of the lowest peak are related
to real and imaginary part of the potential, respectively \cite{Rothkopf:2009pk,Rothkopf:2011db}. 
Motivated by this we calculated Wilson loops on finite temperature lattices in 2+1 flavor QCD using the HISQ action with physical
strange quark mass and light quark masses $m_l=m_s/20$. We performed calculations using $48^3 \times 16$ and $48^3 \times 12$
at $\beta=7.5$ as well as $48^3 \times 12$ lattices at $\beta=7.28$ ($\beta=10/g^2$). These correspond to temperatures $225$ MeV, $300$ MeV and
$249$ MeV respectively. In addition we performed calculations using $24^3 \times 6$ lattices for the lattice parameters
discussed in the previous section. One of the problems in extracting physical information from
the  Wilson loops on the lattice is large noise associated with
them. To reduce the noise smeared gauge fields are used in the spatial gauge transporters $U(\vec{x},\vec{y};\tau)$ 
that enter the Wilson loops.
Alternatively one can fix the Coulomb gauge and omit the spatial gauge connections, i.e calculate correlation function
of Wilson lines of extent $t < 1/T$. This method was used by the MILC collaboration \cite{Aubin:2004wf} as well
as by the HotQCD collaboration \cite{Bazavov:2011nk} to calculate the static potential at zero temperature. We used both approaches.
If the Wilson loop is dominated by the ground state for some value of $\tau$ we may try to extract the static energies 
at non-zero temperature from single exponential fits or from the ratio of the Wilson loops at two neighboring
time-slices separated by single lattice spacing $a$
\be
a V_{eff}(r,\tau)=\ln W(r,\tau/a)/W(r, \tau/a+1).
\ee
At zero temperature for sufficiently large $\tau$ the effective potential $V_{eff}(r,\tau)$ should reach a plateau. 
For non-zero temperature the situation is more complicated due to the backward propagating
contribution. Lattice calculations of the Wilson loops at non-zero temperature in SU(3) gauge theory
show exponential decay in $\tau$ but at distance around $\tau T=1$ the Wilson loops 
increase again \cite{Rothkopf:2009pk,Rothkopf:2011db}.
Similar behavior was observed in 2+1 flavor QCD \cite{Bazavov:2012bq}.
While no temporal boundary conditions are imposed on static
quarks the gluon fields are periodic in time and this may give rise to a contribution that propagates
backward in time. Such backward propagating contribution was also observed in the study of
bottomonium spectral function in NRQCD at non-zero temperature \cite{Aarts:2010ek}.
In Fig. \ref{fig:veff}
we show the effective potential calculated on $N_{\tau}=12$ and $16$ lattices as function of the distance $r$ for
different $\tau$. For $N_{\tau}=16$ lattice that corresponds to the temperature of $225$ MeV plateau seems to
be reached for $\tau T \le 1/2$. For these values of $\tau$ the backward propagating contribution is expected
to be small.
However, the statistical errors are very large for $rT>1$. 
For $N_{\tau}=12$ the effective potential does not reach a plateau for $\tau T \le 1/2$. We do not consider larger values of
$\tau$ because of the backward propagation contribution. We attempted to extract the static energy by removing
the backward propagating contribution and fitting the remainder by single exponential. The results are shown
in Fig. \ref{fig:veff} as open symbols and agree quite well with $V_{eff}(r,\tau=5/(12T))$. Thus it is
reasonable to assume that the static energy is well approximated by  $V_{eff}(r,\tau=5/(12T))$ at these temperatures.
For $rT>1$ the static energy is larger than the free energy. These findings are in agreement with earlier
findings based on $24^3 \times 6$ lattices \cite{Bazavov:2012bq}.

The above analysis as well as the analysis performed in Ref. \cite{Bazavov:2012bq} is based on using the Coulomb gauge. It is
important to check how the results depend on the choice of the static meson operator. In addition it is interesting
to study the onset of medium effects as function of $\tau$. We calculated rectangular Wilson loops using smeared
gauge fields in the spatial gauge transporters. To reduce the noise we used several iterations of APE \cite{Albanese:1987ds}
and HYP \cite{Hasenfratz:2001hp} smearings. 
Namely, we used $5$, $10$ and $20$ steps of APE smearing and $1$, ~$2$ and $5$ steps of HYP smearing.
The numerical results are shown in Fig. \ref{fig:veff_t}.
As expected the HYP smearing is more efficient than APE smearing but for the coarse lattices used in our study
the difference is not that large. One needs 5 steps of HYP smearing or 10 steps of APE smearing to get results 
comparable to the Coulomb gauge results. Fig. \ref{fig:veff_t} shows that except for the lowest temperature and distances
$r T \le 1/2$ the static potential is affected by the medium. While $V_{eff}$ seems to reach a plateau at zero temperature
no plateau is observed at finite temperature. Overall, the behavior of the Wilson loops with smeared spatial links is similar
to the Coulomb gauge result if sufficient number of smearing steps is used. 

\section{Quarkonium spectral functions}
Heavy meson correlation functions in Euclidean time $G(\tau,T)$ are related to the meson spectral functions
$\sigma(\omega,T)$
\begin{equation}
G(\tau,T)=\int_0^{\infty} d \omega \sigma(\omega,T) \frac{\cosh(\omega (\tau-1/(2T)))}{\sinh(\omega/(2T))}.
\end{equation}
Attempts to reconstruct quarkonium spectral function from lattice QCD using the above equation and Maximum
Entropy Method have been presented in Ref. \cite{Umeda:2002vr,Asakawa:2003re,Datta:2003ww}. In these studies it 
was concluded that charmonium ground state may survive in the deconfined medium up to the temperature $1.6$
times the transition temperature or maybe even higher contrary to the expectations based on color screening.
However, the reconstruction of meson spectral functions from Euclidean correlator is very difficult 
\cite{Wetzorke:2001dk}, and the spectral functions are also strongly modified by cutoff effects 
\cite{Karsch:2003wy}. The suggested survival of quarkonium states in the deconfined medium is closely
related to the weak temperature dependence of the Euclidean time quarkonium correlators \cite{Petreczky:2008px}.
Using potential models,  quarkonium spectral functions have been calculated and it was shown that Euclidean correlation
functions do not show significant temperature dependence even if bound states are dissolved 
due to the limited Euclidean time extent \cite{Mocsy:2007yj}.
This seems to be confirmed by the study of spatial charmonium correlators
which indicate the dissolution of the ground state for $T>300$ MeV \cite{Karsch:2012na} as well as by the study of
P-wave bottomonium correlators using lattice NRQCD \cite{Aarts:2010ek}, where larger values of the Euclidean time
can be used.

Potential models can be related to pNRQCD. As the temperature increases the binding energy becomes smaller
and eventual,ly will be the smallest scale in the problem: $E_{bin} \ll \Lambda_{QCD} \ll T,~m_D,~1/r$,
and all the other scales in the problem can be integrated out \cite{Petreczky:2010tk}. 
In this case the potential will be equal to the
static energy. The real part of the static energy can be estimated using lattice QCD. 
In Ref. \cite{Petreczky:2010tk} a phenomenological form based on lattice QCD calculations of the singlet
free energy was used for the real part of the potential while for the imaginary part of the potential 
the hard thermal loop result \cite{Laine:2006ns} was used. The spectral functions calculated in this approach
show that most quarkonia states melt at temperatures $T>250 $MeV, while ground state bottomonium melts at
temperature $T>450$ MeV \cite{Petreczky:2010tk}. The estimates of the static energy obtained from Wilson loops and
discussed in the previous section turn out to be very close to the phenomenological potential used in Ref. \cite{Petreczky:2010tk}.
Therefore the above estimates of the  maximal temperatures that permit the existence of quarkonium states still hold. 

\section{Conclusion}
Color electric screening in high-temperature QCD can be studied using correlation functions
of static mesons that go around the periodic Euclidean time direction. These are related to the
free energy of static quark anti-quark pair. Determination of quarkonium spectral functions from
the meson correlation functions calculated on the lattice is very difficult. The study of Wilson
loops at non-zero temperature offers the possibility to extract the potential that can be used in potential
model calculations to extract the quarkonium spectral functions.

\vspace*{-0.3truecm}
\section*{Acknowledgements}
This work was supported by U.S. Department of Energy under
Contract No. DE-AC02-98CH10886.  Computations have been performed on BlueGene/L computers 
of the New York Center for Computational
Sciences (NYCCS) at Brookhaven National Laboratory
and on clusters of the USQCD collaboration in JLab and FNAL.
\vspace*{-0.3truecm}

\section*{References}
\bibliographystyle{iopart-num}
\bibliography{HotQCD}

\end{document}